\documentclass[journal=jctcce,manuscript=article,layout=traditional]{achemso}
\usepackage{fullpage}
\usepackage{mathtools}
\usepackage{amsthm}
\usepackage{amssymb}
\usepackage{bm}
\usepackage[version=3]{mhchem}
\usepackage{braket}
\usepackage{relsize}
\usepackage{enumitem}
\usepackage[font=small,labelfont=bf]{caption}
\usepackage{graphicx}
\usepackage{footnote}
\usepackage{multirow}
\usepackage{tabularx}
\usepackage{booktabs}
\usepackage{makecell}
\usepackage{threeparttable}
\usepackage{chemfig}

\usepackage{natbib}

\newcommand{\RNum}[1]{\expandafter{\romannumeral #1\relax}}

\newcolumntype{C}{>{\centering\arraybackslash}X}

\title{Revisiting the performance of time-dependent density functional theory for electronic excitations:
Assessment of 43 popular and recently developed functionals from rungs one to four.
}	

\author{Jiashu Liang}
\affiliation{
	Kenneth S. Pitzer Center for Theoretical Chemistry,
	Department of Chemistry,
	University of California at Berkeley,
	Berkeley, CA 94720, USA
}
\author{Xintian Feng}
\affiliation{
	Q-Chem Inc.,
	Pleasanton, CA 94588, USA
}
\author{Diptarka Hait}
\affiliation{
	Kenneth S. Pitzer Center for Theoretical Chemistry,
	Department of Chemistry,
	University of California at Berkeley,
	Berkeley, CA 94720, USA
}
\alsoaffiliation{
	Chemical Sciences Division,
	Lawrence Berkeley National Laboratory,
	Berkeley, CA 94720, USA
}
\author{Martin Head-Gordon}
\affiliation{
	Kenneth S. Pitzer Center for Theoretical Chemistry,
	Department of Chemistry,
	University of California at Berkeley,
	Berkeley, CA 94720, USA
}
\alsoaffiliation{
	Chemical Sciences Division,
	Lawrence Berkeley National Laboratory,
	Berkeley, CA 94720, USA
}
\email{mhg@cchem.berkeley.edu}

\date{\today}

\begin{document}

\begin{tocentry}
\includegraphics[width=5.6cm, height=4 cm]{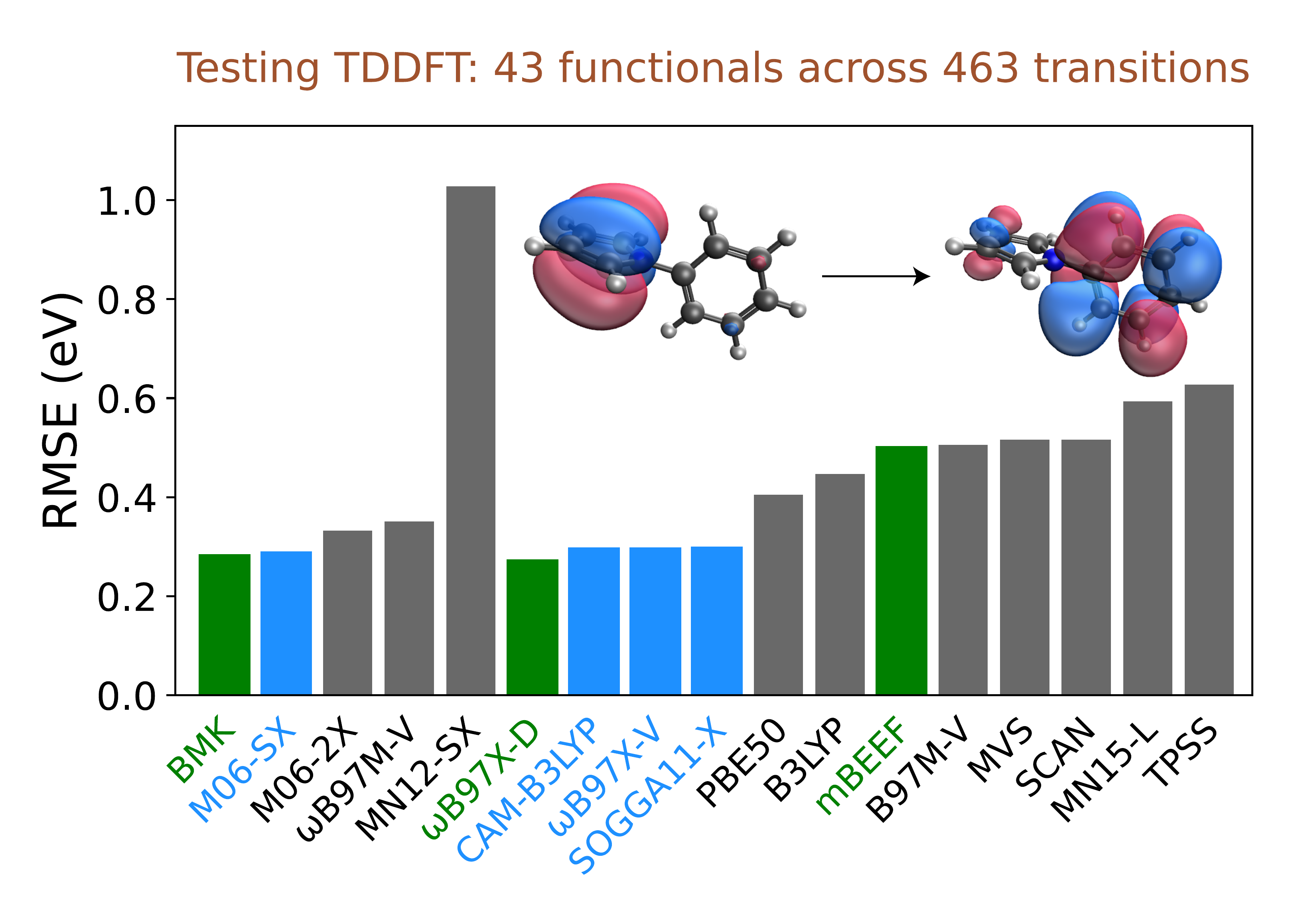}
\end{tocentry}

\begin{abstract}

In this paper, the performance of more than 40 popular or recently developed density functionals is assessed for the calculation of 463 vertical excitation energies against the large and accurate QuestDB benchmark set. For this purpose, the Tamm-Dancoff approximation offers a good balance between performance and accuracy. The functionals $\omega$B97X-D and BMK are found to offer the best performance overall with a Root-Mean Square Error (RMSE) of 0.28 eV, better than the computationally more demanding CIS(D) wavefunction method with a RMSE of 0.36 eV. The results also suggest that Jacob's ladder still holds for TDDFT excitation energies, though hybrid meta-GGAs are not generally better than hybrid GGAs. Effects of basis set convergence, gauge invariance correction to meta-GGAs, and nonlocal correlation (VV10) are also studied, and practical basis set recommendations are provided.

\end{abstract}

\maketitle

\clearpage

\section{Introduction} \label{sec:intro}

Time-dependent (TD) density functional theory (DFT) is currently the most popular approach to model the electronic structures and properties of molecular excited states.\cite{dreuw2005single, marques2004time, burke2005time} Although TDDFT is a formally exact theory supported by the Runge-Gross theorem\cite{runge1984density}, it is not practically possible to find the true time-dependent functional that maps electron density to electronic energy. Therefore, approximations have to be introduced. Most practical calculations employ the adiabatic local density approximation (ALDA) in which the time-dependent functional can be replaced by a standard time-independent one (in the limit of the density slowly varying in time). The ALDA allows TDDFT to directly employ various density functional approximations designed for ground-state DFT calculations, without the need for any additional parameters. However, as consequences of these approximations and the linear response (LR) formalism\cite{casida1995recent} adopted, the performance of a specific functional in LR-TDDFT calculations cannot be gauged a priori and instead must be assessed against benchmark excitation energies.

Since the 2000s, several benchmark studies have explored the performance of many popular functionals against experimental\cite{jacquemin2009extensive, caricato2010electronic, leang2012benchmarking, suellen2019cross} or theoretical reference transition energies \cite{silva2008benchmarks, sauer2009benchmarks, silva2010basis, jacquemin2011excited, goerigk2009computation, jacquemin2009extensive, peverati2012performance, mardirossian2011benchmark, peach2008excitation, peach2012overcoming, rohrdanz2009long, nguyen2011performance, shukla2010comprehensive, wong2009absorption, guo2012time} within the LR-TDDFT framework. When we compare with experimental values, we need to account for many additional factors (temperature, vibrational structure, vibronic coupling, environmental effects, and so on) and it is not always straightforward to map the computed electronic transitions to the experimental spectra, especially for high-lying excited states. These problems  can be avoided by comparing against computed reference values, at specified molecular geometries. One of the most popular theoretical set is the 28-molecules list developed by Thiel and coworkers in 2008\cite{silva2008benchmarks}. Thiel's set consists of theoretical best estimates (TBE) for 104 singlet and 63 triplet excitations, which were subsequently extended and used by Thiel and many other groups.\cite{sauer2009benchmarks, silva2010basis, jacquemin2011excited, goerigk2009computation, jacquemin2009extensive, peverati2012performance, mardirossian2011benchmark} Another popular set is a list of 59 singlet states developed by Tozer and coworkers, which gives insight into valence (local), Rydberg, and charge-transfer (CT) excitations.\cite{peach2008excitation, peach2012overcoming, mardirossian2011benchmark, rohrdanz2009long, nguyen2011performance} Some other works also involved a small number of compounds.\cite{zwijnenburg2008optical, wong2008coumarin, wong2009absorption, shukla2010comprehensive, guo2012time, list2012performance}  For a detailed review on the topic, see Ref.~\citenum{laurent2013td}. 

However, the theoretical datasets used by the previous benchmarks also suffer from some problems due to  computational limitations. Some datasets are quite small with only a few Rydberg and CT states\cite{silva2008benchmarks, jacquemin2011excited, goerigk2009computation}, or lack singlet-triplet excitations.\cite{jacquemin2009extensive, goerigk2009computation} Some datasets' reference methods, like CC2\cite{christiansen1995second, shukla2010comprehensive, wong2009absorption}, are not substantially more accurate than TDDFT. Previous benchmark studies\cite{jacquemin2011excited, peach2012overcoming} show that the best functionals for LR-TDDFT can have a mean absolute error (MAE) around 0.3 eV. Therefore, to assess the best functionals, the reference wavefunction theory is required to have an MAE less than 0.1 eV, like CC3\cite{christiansen1995response} and other methods containing triple excitations. The dearth of comprehensive and reliable datasets has led recent benchmarks to only focus on specific molecules\cite{zhou2018lowest, shao2019benchmarking, lopez2020ab} or specific excitation types\cite{li2016critical, bussy2021efficient}. The recent development of the accurate and comprehensive QUEST database,\cite{loos2018mountaineering, loos2020mountaineering1, loos2020mountaineering2, veril2021questdb, loos2021reference, loos2021mountaineering} however, allows for the re-examination of the accuracy of LR-TDDFT.  We choose 400 TBE of the vertical excitations from the dataset. Most of them are claimed to be within 0.05 eV (or less) of the full configuration interaction (FCI) limit, which is much more accurate than the previous datasets. Additionally, QUEST TBE values also account for the most important one-particle basis set effects, which is necessary to enable comparison between wavefunction-based reference values and LR-TDDFT values close to the complete basis set limit.

One further consideration that makes revisiting the performance of LR-TDDFT timely is the fact that many new functionals have been developed in the recent 15 years but have not been benchmarked for predicting excitation energies. For example, our group has developed three combinatorically optimized semi-empirical functionals: $\omega$B97X-V\cite{Mardirossian:2014}, a range-separated hybrid (RSH) generalized-gradient approximation (GGA), B97M-V\cite{Mardirossian:2015}, a local meta-GGA (mGGA), and $\omega$B97M-V\cite{Mardirossian:2016}, an RSH mGGA. They are independently parameterized, but all employ the VV10 nonlocal correlation functional\cite{vydrov2010nonlocal} (NLC) to correctly predict long-range dispersion interactions. These functionals are notably more accurate compared to other functionals for calculating various types of ground-state energies.\cite{mardirossian2017thirty, najibi2018nonlocal, chan2019assessment} But there is no TDDFT benchmark involving them so far, possibly due to the difficulty of deriving and coding the second derivative formula of nonlocal functionals.\cite{miwa2022linear} Besides the B97 family, many new functionals also have been developed by other groups and show very good performance for predicting ground-state properties. We choose MVS family\cite{sun2015semilocal}, MS2 family\cite{sun2013semilocal}, and M06-SX functional\cite{wang2020m06} as the examples. To the best of our knowledge, this paper is the first time that their performance for predicting the excitation energies has been assessed.

The remainder of this paper is organized as follows. We first briefly review the theory of LR-TDDFT, the  Tamm–Dancoff approximation\cite{hirata1999time1} (TDA) and the mGGA gauge invariance (GINV) correction\cite{bates2012harnessing} in Section~\ref{sec:method} and then describe the computational details in Section~\ref{sec:comp}. After the discussion of the GINV correction, VV10's effect and the basis set convergence using the highly accurate QUEST $\#$1 dataset of small molecules (Section~\ref {sec:bench_method}), the performance of 43 functionals for predicting excitation energy is assessed by a more comprehensive benchmark set (Section~\ref{sec:bench_functional}). Concluding remarks are given in Section~\ref{sec:conclusion}.

\section{Method} \label{sec:method}

\subsection{Linear response time-dependent density functional theory and Tamm–Dancoff approximation} \label{subsec:lrtddft_tda}

 Time-dependent density functional theory in the linear-response formalism employs the following non-Hermitian eigenvalue equation to calculate the excitation energy $\omega$. 

\begin{equation}
        \begin{bmatrix} \mathbf{A} &\mathbf{B} \\\mathbf{B^*}&\mathbf{A^*} \end{bmatrix}
                \begin{bmatrix} \mathbf{X} \\ \mathbf{Y} \end{bmatrix} 
                = \omega \begin{bmatrix} \mathbf{1} &\mathbf{0} \\\mathbf{0}&\mathbf{-1} \end{bmatrix}
                \begin{bmatrix} \mathbf{X} \\ \mathbf{Y} \end{bmatrix}
\end{equation}
Here, $\mathbf{X}$ and $\mathbf{Y}$ are the transition vectors. The elements of the matrices $\mathbf{A}$ and $\mathbf{B}$ are given as

\begin{equation}
        {A}_{ai,bj} = \delta_{ij}\delta_{ab}(\epsilon_a - \epsilon_i) + (ai|jb) - c_{\text{HF}}(ab|ji) + (1 -c_{\text{HF}})(ai|f_{xc}|jb)
\end{equation}
\begin{equation}
        {B}_{ai,bj} = (ai|bj) - c_{\text{HF}}(bi|aj) + (1 -c_{\text{HF}})(ai|f_{xc}|bj)
\end{equation}
where the indices $i$, $j$ and $a$, $b$ label occupied and virtual orbitals respectively, which will also be written as $\psi_i$, $\psi_j$, $\psi_a$ and $\psi_b$ later. We will use $m$ and $n$ to represent arbitrary molecular orbitals, while $\epsilon_a$ and $\epsilon_i$ are orbital energies of Kohn-Sham orbitals $a$ and $i$. $(ai|jb)$ denotes an electron repulsion integral in Mulliken notation (Equation~\ref{Mulliken}) and $f_{xc}$ is the exchange-correlation (xc) potential. In the adiabatic local density approximation, the response of the xc potential corresponds to the second functional derivative of the xc energy, given as
\begin{equation}\label{Mulliken}
        (ai|jb) = \int d^3rd^3r'\, \phi_a^*(r) \phi_i(r) \frac{1}{|r-r'|} \phi_j^*(r')\phi_b(r') 
\end{equation}
\begin{equation}
        (ai|f_{xc}|jb) = \int d^3rd^3r'\, \phi_a^*(r) \phi_i(r)\frac{\delta^2E_{xc}}{\delta\rho(r)\delta\rho(r')} \phi_j^*(r')\phi_b(r') 
\end{equation}

In the Tamm–Dancoff approximation, the \textbf{B} matrix is neglected, and the excitations are decoupled from the de-excitations, yielding:
\begin{equation}
        \mathbf{A}\mathbf{X} = \omega\mathbf{X}
\end{equation}
TDA does not obey the Thomas-Reiche-Kuhn sum rule, which states that the
sum of oscillator strengths must equal the number of electrons. Therefore, the calculated TDA transition moments cannot be expected to be more than qualitatively accurate. However, the TDA eigenvalue equation is analogous to variational CIS, resulting more resistance to the triplet instabilities\cite{peach2013triplet}.

Efficient Davidson-type algorithms are always used in the actual implementation of both TDDFT and TDA, to iteratively obtain a small number of lowest excitation energies and corresponding transition vectors.\cite{davidson1975, stratmann1998efficient} These algorithms need to evaluate generalized matrix-vector products (i.e. the contraction of a $4^\mathrm{th}$ tensor such as $\mathbf{A}$ with a $2^\mathrm{nd}$ rank tensor such as $\mathbf{X}$) to yield another matrix, $\mathbf{G=AX}$ on each iteration. The xc part of $\mathbf{G}$ can be evaluated as
\begin{equation}\label{Ax}
\begin{split}
        G_{ai} = \sum_{jb}(ai|f_{xc}|jb) X_{bj} &= \sum_{jb}\sum_{\mu\nu\lambda\kappa} C_{\nu a}^* C_{\mu i} (\nu\mu|f_{xc}|\lambda\kappa) C_{\lambda j}^* C_{\kappa b} X_{bj} \\
        &= \sum_{\mu\nu}C_{\nu a}^* C_{\mu i} \sum_{\lambda\kappa}\frac{\partial^2E_{xc}}{\partial {P}_{\mu\nu}\partial {P}_{\kappa\lambda}} \sum_{jb} C_{\lambda j}^* C_{\kappa b} X_{bj}
\end{split}
\end{equation}
Here, $\mu$, $\nu$, $\lambda$ and $\kappa$ are atomic orbitals, which can also be written as $\chi_{\mu}$, $\chi_{\nu}$, $\chi_{\lambda}$ and $\chi_{\kappa}$. $C_{\mu i}$, $C_{\nu a}$, $C_{\lambda j}$, $C_{\kappa b}$ are molecular orbital coefficients. We can form a density-matrix-like quantity (we can call it a trial density matrix\cite{hirata1999configuration}) defined by
\begin{equation}
        {P}_{\kappa\lambda}^t = \sum_{jb}  C_{\kappa b} X_{bj} C_{\lambda j}^*
\end{equation}
And thus we can calculate the following Fock-matrix-like quantity and substitute it back into Equation~\ref{Ax}.
\begin{equation}\label{Gxc}
        {G}_{\nu\mu}^t = \sum_{\lambda\kappa} \frac{\partial^2E_{xc}}{\partial {P}_{\mu\nu}\partial {P}_{\lambda\kappa}} {P}_{\kappa\lambda}^t
\end{equation}
\begin{equation}
        G_{ai} = \sum_{jb}(ai|f_{xc}|jb) X_{bj} =\sum_{\mu\nu}C_{\nu a}^* C_{\mu i} {G}_{\nu\mu}^t 
\end{equation}

\subsection{Gauge invariance correction to mGGA} 

The kinetic energy density of mGGA functionals is usually defined as
\begin{equation}
        \tau(r, t) =\frac{1}{2} \sum_{j}|-i\nabla \psi_j(r, t) |^2
\end{equation}
which changes upon a gauge transformation (i.e. $\tau(r, t)$ is gauge-dependent), leading to the energy also being gauge-dependent. This can be shown using the real gauge function $\Lambda(r,t)$.
\begin{equation}
        \psi_j[\Lambda](r, t) = \psi_j(r, t)e^{-i\Lambda(r,t)}
\end{equation}
\begin{equation}
        \tau[\Lambda](r, t) = \tau(r, t) - \nabla\Lambda(r,t) \cdot \mathbf{j}_p(r,t) + \frac{1}{2}|\nabla\Lambda(r,t)|^2 \rho(r,t)
\end{equation}
Here $\mathbf{j}_p(r,t)$ is the paramagnetic orbital current density of the original system.
\begin{equation}
\begin{split}
        \mathbf{j}_p(r,t) &= \sum_{j} -i\psi_j^*(r, t)\nabla\psi_j(r, t) + i\psi_j(r, t)\nabla\psi_j^*(r, t) \\
        &= \sum_{\mu\nu} i {P}_{\mu\nu}(\chi_{\mu}\nabla\chi_{\nu} - \chi_{\nu}\nabla\chi_{\mu})
\end{split}
\end{equation}
$\rho(r,t)$ is the total electron density.

For current-free ground states, the gauge may be fixed by choosing real KS orbitals and thus the mGGA is perfectly well defined. However, for time-dependent and current-carrying states, this becomes a problem. Recently, Bates and Furche\cite{bates2012harnessing} developed a corrected kinetic energy density $\hat{\tau}$, which is gauge invariant and satisfies the iso-orbital constraint.
\begin{equation}
        \hat{\tau}(r,t) =  \tau(r, t) -\frac{|\mathbf{j}_p(r,t)|^2}{2\rho(r,t)}
\end{equation}

In the case of a static ground state, $\mathbf{j}_p(r,t)$ will vanish and $\hat{\tau}$  will reduce to the original $\tau$ since the density matrix ${P}_{\mu\nu}$ is real and symmetric. In other words, the correction has no contribution to the Fock matrix and thus has no influence on the ground-state self-consistent field (SCF) procedure. But it will change the TDDFT result since the trial density, $\mathbf{P}^t$, is not symmetric. One more term needs to be added to $\mathbf{G}^t$, given by:
\begin{equation}
\begin{split}
        {G}_{\nu\mu}^{corr,t} &= - \int d^3\mathbf{r}d^3\mathbf{r}'\, \frac{1}{\rho}\frac{\partial^2 f}{\partial\tau^2}
        \Big\lvert\frac{\partial\mathbf{j}_p}{\partial {P}_{\mu\nu}} \cdot \sum_{\lambda\kappa}\frac{\partial\mathbf{j}_p}{\partial {P}_{\lambda\kappa}} {P}_{\kappa\lambda}^t \Big\rvert\\
        &=- \int d^3\mathbf{r}d^3\mathbf{r}'\, \frac{1}{\rho}\frac{\partial^2 f}{\partial\tau^2} \Big\lvert(\chi_{\mu}\nabla\chi_{\nu} - \chi_{\nu}  \nabla\chi_{\mu}) \cdot \mathbf{j}_p^t \Big\rvert
\end{split}
\end{equation}
where
\begin{equation}
        \mathbf{j}_p^t =  \sum_{\lambda\kappa} (\chi_{\lambda}\nabla\chi_{\kappa} - \chi_{\kappa}\nabla\chi_{\lambda}) {P}_{\kappa\lambda}^t
\end{equation}
Here we have applied the condition of $|\mathbf{j}_p| = 0$ to avoid the cross terms between $\tau$ and other density variables. Additionally we have adopted the approximation $\frac{\partial^2 f}{\partial\hat{\tau}^2} = \frac{\partial^2 f}{\partial\tau^2}$ to employ existing functionals.

\section{Computational Details} \label{sec:comp}

We have tested the following density functionals: 
\begin{itemize}[label={$\bullet$}]
    \item Local Density Approximation (LDA, Rung 1): SPW92\cite{dirac1930note,perdew1992accurate}
    \item GGA (Rung 2): B97-D\cite{grimme2006semiempirical}, MPW91\cite{adamo1998exchange, perdew1992atoms}, PBE\cite{perdew1997generalized}, BLYP\cite{miehlich1989results}, N12\cite{peverati2012improved}, SOGGA11\cite{peverati2011generalized}.
    \item mGGA (Rung 3): B97M-V\cite{Mardirossian:2015}, mBEEF\cite{wellendorff2014mbeef}, SCAN\cite{sun2015strongly}, MS2\cite{sun2013semilocal}, M06-L\cite{zhao2006new}, MVS\cite{sun2015semilocal}, revM06-L\cite{wang2017revised}, MN15-L\cite{yu2016mn15}, revTPSS\cite{perdew2009workhorse}, TPSS\cite{tao2003climbing}.
    \item hybrid GGA (Rung 4): $\omega$B97X-D\cite{chai2008long}, CAM-B3LYP\cite{yanai2004new}, $\omega$B97X-V\cite{Mardirossian:2014}, SOGGA11-X\cite{peverati2011communication}, LRC-wPBE\cite{rohrdanz2008simultaneous}, LRC-wPBEh\cite{rohrdanz2009long}, MPW1K\cite{lynch2000adiabatic}, PBE0\cite{adamo1999toward}, HSEHJS\cite{henderson2008generalized,krukau2006influence}, rcamB3LYP\cite{cohen2007development}, MPW1PW91\cite{adamo1998exchange}, BHHLYP\cite{Becke1988,Lee1988}, PBE50\cite{bernard2012general}, B3LYP\cite{Becke1993,stephens1994ab}, HFLYP\cite{Lee1988}.
    \item hybrid mGGA (Rung 4): BMK\cite{boese2004development}, M06-SX\cite{wang2020m06}, M06-2X\cite{zhao2008m06}, $\omega$B97M-V\cite{Mardirossian:2016}, wM05-D\cite{lin2012long}, MN15\cite{haoyu2016mn15}, PW6B95\cite{zhao2005design}, SCAN0\cite{hui2016scan}, MS2h\cite{sun2013semilocal}, M11\cite{peverati2011improving}, revTPSSh\cite{csonka2010global}, TPSSh\cite{staroverov2003comparative}, MVSh\cite{sun2015semilocal}, MN12-SX\cite{peverati2012screened}.
\end{itemize}

For comparison, we also tested two standard wavefunction methods: Configuration interaction singles (CIS)\cite{foresman1992toward} and CIS with perturbative double corrections [CIS(D)]\cite{head1994doubles}. We do not benchmark double hybrid functionals in this paper, but interested readers can consult Ref.~\citenum{goerigk2009computation} and Refs.~\citenum{grimme2007double, casanova2019omega, casanova2020assessing, casanova2021time}.

All the calculations are performed using a development version of Q-Chem 5.4\cite{epifanovsky2021software}. Molecular geometries employed are directly obtained from the QUEST database\cite{veril2021questdb}. Local xc integrals are calculated over a radial grid with 99 points and an angular Lebedev grid with 590 points for all atoms, while non-local VV10 correlation is calculated over an SG-1\cite{gill1993} grid (which is a subset of points employed in a grid with 50 radial and 194 angular points). Section~\ref{subsec:mgga_ginv} indicates that use of the TDA is preferable, and all results beyond that point utilize it (unless specified otherwise). 

We select reference transitions from QUEST database according to the following three criteria: ($\rm \RNum{1}$) the transition should be labelled ``safe'' (the deviation with FCI is expected to be smaller than 0.05 eV) ($\rm \RNum{2}$) the transition should have dominant single-excitation character ($\%\rm T_1 > 85\,\%$, as computed at the CC3 level) except for the data in the QR dataset ($\rm \RNum{3}$) the electronic state associated with the transition should be assignable using the attribution procedure described below.

All the calculations employ the aug-cc-pVTZ basis\cite{dunning1989gaussian, kendall1992electron, woon1993gaussian} except for the basis set tests described in Section~\ref{subsec:basis}. The large aug-cc-pVTZ basis set contains many diffuse basis functions and thus yields significant orbital mixing in the excitation amplitudes, making the attribution of the various transitions very challenging. We calculate three times the number of excitations in the benchmark dataset for each molecule, and then divide them into different categories according to their symmetry and spin. For each target state, we search in its corresponding category and find the lowest-energy state which contains the required orbital excitation type (e.g. $\pi\to\pi^*$). Then we calculated the root mean square electron size of the excited state, $\sigma_e = (\langle\vec{x}_e^{\,2}\rangle - \langle\vec{x}_e\rangle^2)^{1/2}$ through the one-electron transition-density matrix\cite{Plasser2014}. If the size satisfies the valence/Rydberg type required ($\sigma_e < 2.5$ \AA\, for valence states and $\sigma_e > 2.5$ \AA\, for Rydberg states), we will attribute this state to the target state. If not, we will compare all the excited states in this category to choose the most suitable state. We also manually checked states that exhibit  big differences with the reference energy ($>\,$1.0 eV) in case there are exceptions to the assignment criteria employed above, such as some valence states which has a very big $\sigma_e$. 

The selected molecules are divided into five subsets: Q1, Q2, QE, QR and QCT. The molecules in Q1 are from QUEST $\#$1 subset\cite{loos2018mountaineering}. The reference accuracy of this subset reaches FCI level but it only contains molecules with 4 non-hydrogen atoms at most. Q2 are adapted from QUEST $\#$3\cite{loos2020mountaineering1} and $\#$5\cite{veril2021questdb} , which have slightly lower benchmark accuracy but contains molecules with more non-hydrogen atoms (10 at most). QE and QR are obtained from the ``Exotic'' and ``Radical'' subsets of QUEST $\#4$\cite{loos2020mountaineering2}, while QCT is QUEST $\#$6 dataset\cite{loos2021reference}. The numbers of the different types of states in each dataset is summarized in Table~\ref{tab:states_stat}. Only the Q1 subset is used in Section~\ref{sec:bench_method} if not otherwise stated.

\begin{table}[ht!]
\caption{The number of different types of the excited states in each dataset}
\label{tab:states_stat}
\begin{tabular}{l|cccccc}
        & Q1  & Q2  & QE & QCT                                                                      & QR                                                                      & all \\ \hline
Singlet & 55  & 154 & 19 & \multirow{4}{*}{\begin{tabular}[c]{@{}c@{}}All\\  Singlets\end{tabular}} & \multirow{4}{*}{\begin{tabular}[c]{@{}c@{}}All\\ Doublets\end{tabular}} & 254 \\
Triplet & 45  & 111 & 11 &                                                                          &                                                                         & 167 \\
Valence & 60  & 201 & 28 &                                                                          &                                                                         & 289 \\
Rydberg & 39  & 64  & 2  &                                                                          &                                                                         & 105 \\ \hline
Total   & 100 & 265 & 30 & 27                                                                      & 42                                                                      & 463\\ 
\hline  
\end{tabular}
\end{table}

\section{Preliminary benchmark of methods and basis sets} \label{sec:bench_method}

\subsection{Effect of TDA approximation and meta-GGA gauge invariance correction}\label{subsec:mgga_ginv}

(This section has been rewritten after the review. We find the difference between TDDFT and TDDFT/TDA is mainly on triplet valence subset and it's state-specific. For further detatils, please see published paper: https://doi.org/10.1021/acs.jctc.2c00160)

\begin{figure}[ht]
    \centering
    \includegraphics[width=0.95\textwidth]{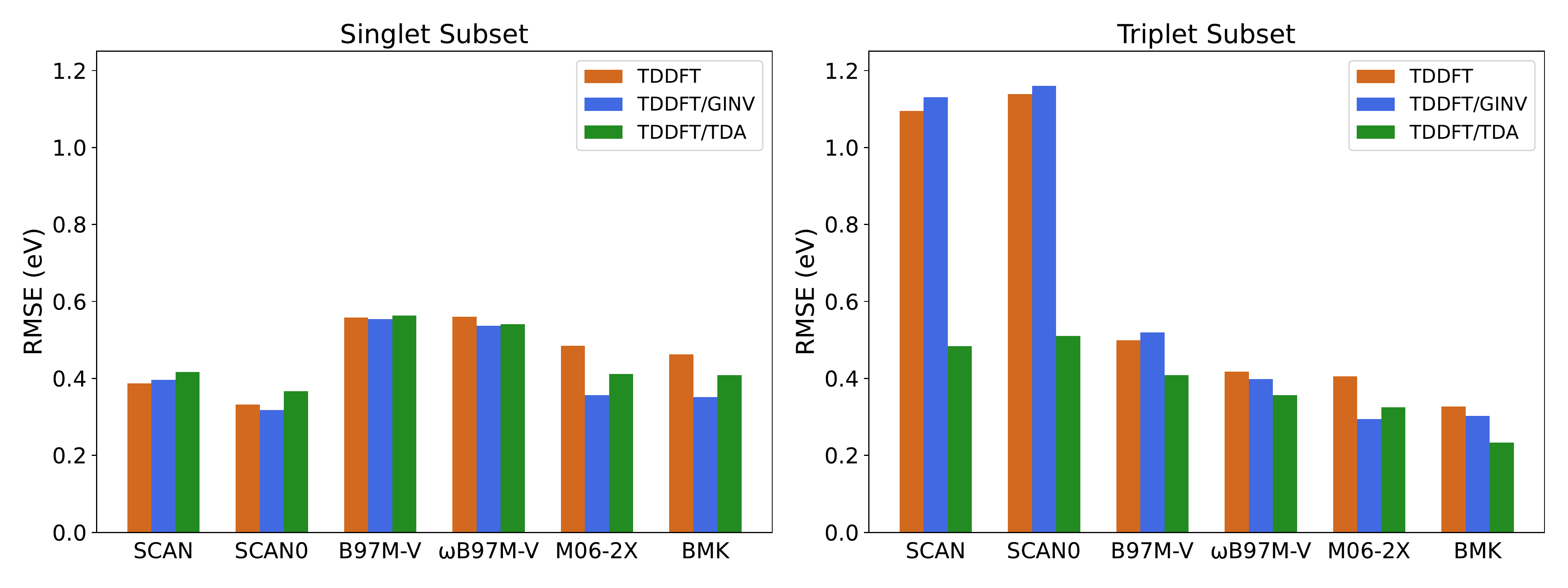}
    \caption{Comparison of the Root-Mean Square Errors (RMSE) of TDDFT, TDDFT/GINV, TDDFT/TDA for singlet and triplet subsets}
    \label{fig:TD}
\end{figure}

We first explore the effect of TDA and GINV correction on the mGGA and hybrid mGGA functionals separately. The GINV correction is only suitable for use with TDDFT because TDA itself breaks gauge invariance.\cite{furche2001density} Figure~\ref{fig:TD} shows graphical representations of the Root-Mean Square Error (RMSE) of six representative functionals for singlet and triplet excited states via TDDFT, TDDFT with GINV correction, and TDA. The results obtained with other tested functionals (i.e. mBEEF, M06-L, MS2 and MS2h) are summarized in Table~S3.1.

For the singlet states subset, the RMSE of the three TDDFT approaches varies only slightly, with differences smaller than 0.03 eV for most tested functionals. The only exceptions are the interesting cases of M06-2X and BMK, where the RMSE associated with TDDFT is reduced by a quite significant 0.1 eV by the GINV correction. For triplet excited states, the effect of GINV correction is still quite small for 10 of the 11 tested functionals, but TDA substantially decreases the RMSE associated with TDDFT for all functionals. The M06-2X functional stands out again: the use of TDDFT with the GINV correction yields a slightly lower RMSE than the TDA.

From the raw data in Table~S1.1, we can see that TDDFT/TDA can significantly improve the accuracy of calculated triplet excitations when pure TDDFT suffers an instability problem, as happens for the $^3\Sigma_u^+$  state of nitrogen, $^3\Sigma^+$ state of carbon monoxide and so on. This advantage of TDA has been pointed by many previous studies.\cite{hirata1999time1, peach2013triplet, peach2012overcoming, laurent2013td} For most functionals it seems that the TDDFT/GINV method still inherits the problem from pure TDDFT. Overall, TDDFT/TDA is recommended for most cases since it is computationally more efficient than full TDDFT and can improve the prediction of triplet transitions. We will adopt this method for all calculations reported in the following sections.

\subsection{Basis set convergence}\label{subsec:basis}

\begin{figure}[ht!]
    \centering
    \includegraphics[width=0.95\textwidth]{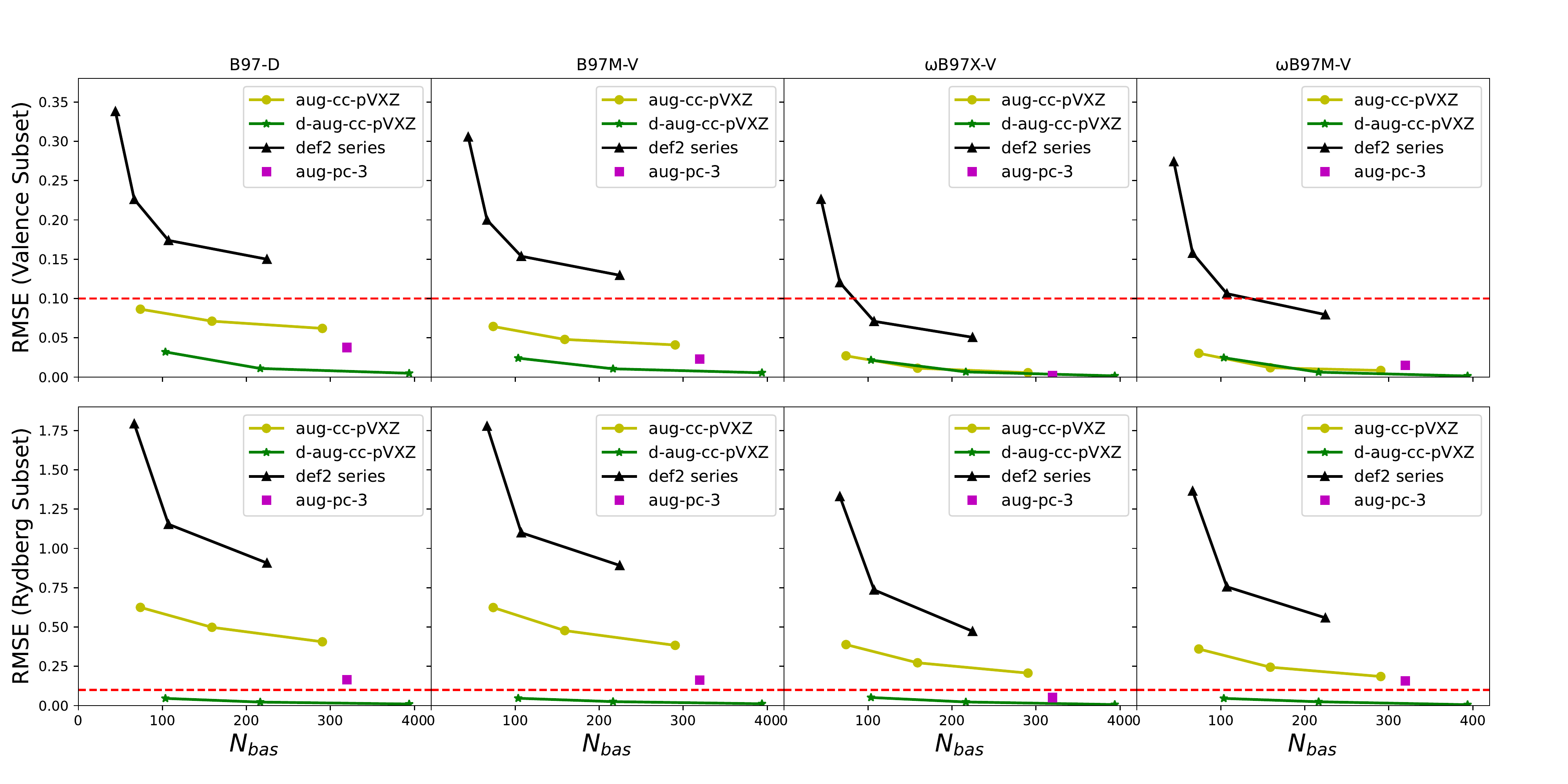}
    \caption{Comparison of the RMSEs (in eV) of several basis sets compared to d-aug-cc-pV5Z for valence and Rydberg subsets of Q1. Tested basis sets includes: aug-cc-pVDZ, aug-cc-pVTZ, aug-cc-pVQZ;\cite{dunning1989gaussian, kendall1992electron, woon1993gaussian} d-aug-cc-pVTZ, d-aug-cc-pVQZ;\cite{woon1994gaussian} def2-SVP (only for valence subset), def2-SVPD, def2-TZVPD, def2-QZVPPD;\cite{weigend2005balanced, rappoport2010property} and aug-pc-3\cite{jensen2002polarization}. The red horizontal line indicates an error of 0.1 eV.}
    \label{fig:basis}
\end{figure}

\begin{figure}[ht!]
    \centering
    \includegraphics[width=\textwidth]{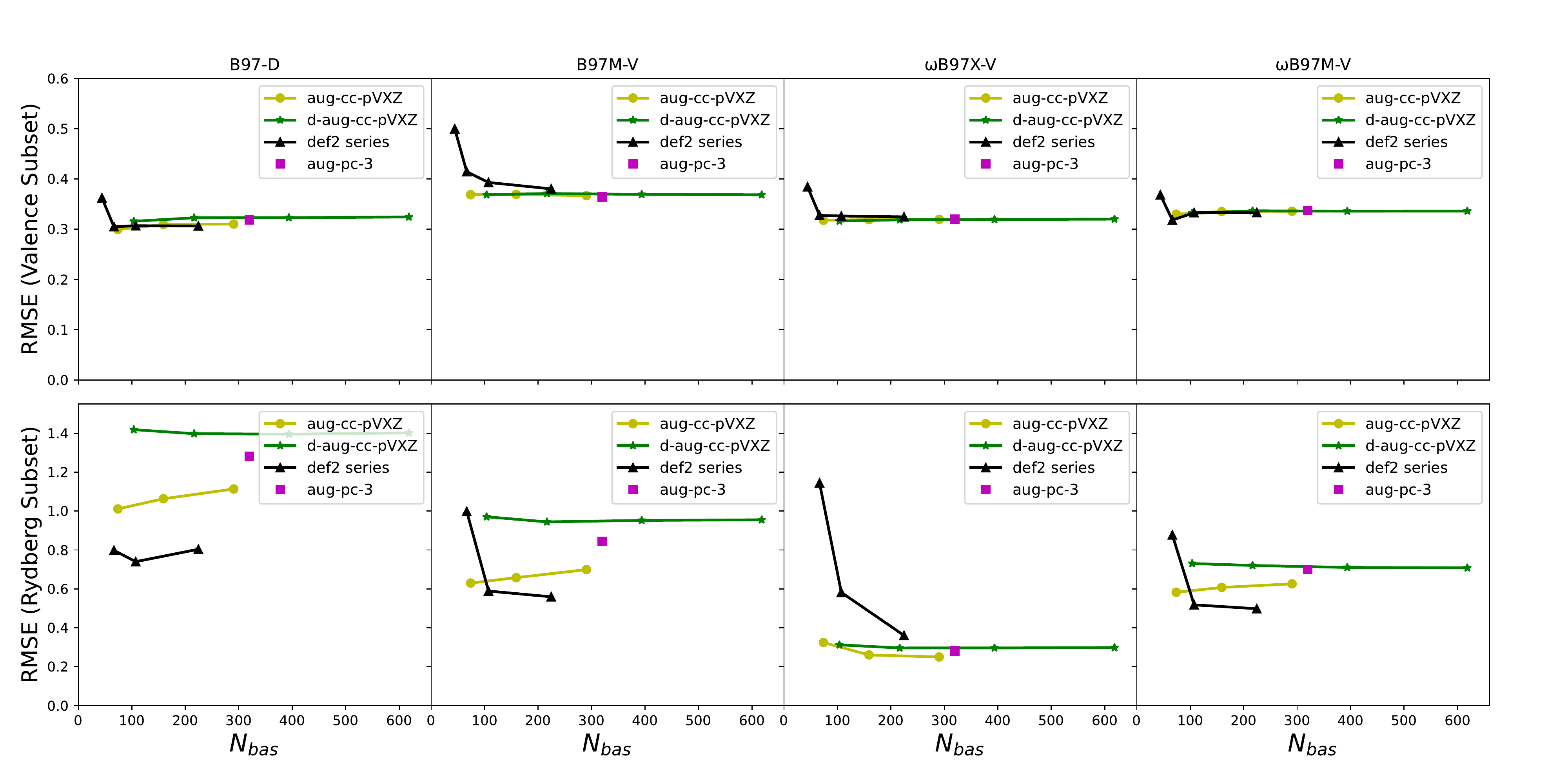}
    \caption{Comparison of the RMSEs (in eV) relative to TBE/CBS for Valence and Rydberg subsets of Q1. Tested basis sets includes aug-cc-pVDZ, aug-cc-pVTZ, aug-cc-pVQZ, d-aug-cc-pVTZ, d-aug-cc-pVQZ, d-aug-cc-pV5Z, def2-SVP (only for Valence subset), def2-SVPD, def2-TZVPD, def2-QZVPPD and aug-pc-3.}
    \label{fig:basis_CI}
\end{figure}

Taking B97-D, B97M-V, $\omega$B97X-V and $\omega$B97M-V as examples, we explore the basis convergence of functionals from different rungs of Jacob's Ladder.\cite{perdew2005prescription} Figure~\ref{fig:basis} and \ref{fig:basis_CI} display the RMSEs obtained with different basis sets for valence and Rydberg subsets against their average number of basis functions across the Q1 dataset. We employ two kind of reference values here. One is calculated by the method itself in the complete basis limit (CBS, here approximated with d-aug-cc-pV5Z basis set\cite{woon1994gaussian}), and the other is the set of TBE/CBS values from Ref.~\citenum{loos2018mountaineering}.  

It is evident that $\omega$B97X-V and $\omega$B97M-V converge more rapidly than B97-D, B97M-V when increasing the basis set size, no matter which basis set family is used. This seems to imply that the hybrid functionals may converge more rapidly than the local functionals. But due to the limited number of tested functionals, we cannot rule out the possibility that the basis set convergence may be just influenced by characteristics of the individual functional. For instance, it has been previously demonstrated that reaching the CBS limit even for ground-state intermolecular interactions is extremely difficult with certain Minnesota density functionals.\cite{mardirossian2013characterizing}

Figure~\ref{fig:basis} also demonstrates that the convergence for Rydberg states is more difficult than that for valence states. The RMSE of the aug-cc-pVDZ basis for the valence subset is already below 0.1 eV, which is only one third of the smallest method error described in Section~\ref{sec:bench_functional}. In contrast, double augmentation with diffuse functions must be employed if we want to achieve similar accuracy for Rydberg states. Interestingly, Figure~\ref{fig:basis_CI} shows that adding more diffuse functions will make predictions worse compared to TBE/CBS. A likely reason is that TDDFT systematically underestimates Rydberg excitation energies and a larger basis set will aggravate this problem. This can be supported by the fact that the basis set with a higher RMSE (vs TBE) always has a more negative mean signed error (MSE). 

Overall, the d-aug-cc basis set family performs best at reaching the CBS limit for most functionals. The aug-cc-pVXZ family appears to be only adequate for valence excited states with hybrid functionals. The def2 family is surprisingly unsuitable for attaining the CBS limit in the cases assessed here. However, as a result, the def2 basis family usually offers good cancellation of basis error and method error for predicting Rydberg excited states. Therefore, for practical purposes, we recommend aug-cc-pVDZ to predict valence states at moderate computational cost. This same basis set is also a reasonable choice for Rydberg states with hybrid density functionals, while def2-TZVPD seems to be a reasonable alternative for Rydberg states due to fortuitous error cancellation.

\subsection{Effect of VV10}

The effect of nonlocal correlation functional VV10 was also studied. The detailed implementation of the VV10 contribution to ${G}_{\nu\mu}^t$ (Equation~\ref{Gxc}) necessary to perform these tests will be presented elsewhere. Table~\ref{tab:vv10} shows that the change of the RMSE and MSE by removing VV10 is smaller than 0.01 eV, suggesting VV10 has very little effect on the prediction of the excitation energies in the Q1 subset. This supports the claim of Najibi and Goerigk that inclusion of nonlocal correlation functionals in SCF has no significant effect on orbital-energy differences.\cite{najibi2018nonlocal} We also compare $\omega$B97X($\omega$B97M) and $\omega$B97X-V($\omega$B97M-V) on the Q2 dataset and their result are also very close (shown in SI), further validating this conclusion.

\begin{table}[ht!]
\caption{Comparison of RMSE and MSE (both in eV) of B97M-V, $\omega$B97X-V and $\omega$B97M-V with or without VV10's contribution. $\omega$B97X in this table refers to $\omega$B97X-V without the VV10 contribution, not the stand-alone functional developed by Chai et al. in 2008\cite{Chai2008}. }
\label{tab:vv10}
\begin{tabular}{l|cc}
\multicolumn{1}{l|}{} & \multicolumn{1}{c}{\textbf{RMSE}} & \multicolumn{1}{c}{\textbf{MSE}} \\ \hline
\textbf{B97M}         & 0.502                            & -0.150                            \\
\textbf{B97M-V}       & 0.498                            & -0.148                           \\ \hline
\textbf{$\omega$B97X$^a$}        & 0.279                            & -0.119                           \\
\textbf{$\omega$B97X-V}      & 0.284                            & -0.109                           \\ \hline
\textbf{$\omega$B97M}        & 0.465                            & -0.329                           \\
\textbf{$\omega$B97M-V}      & 0.464                            & -0.329                        \\ \hline  
\end{tabular} 
\end{table}

\section{Comprehensive benchmark of 43 density functional approximations} \label{sec:bench_functional}

\subsection{Overall performance}\label{subsec:overall}

\begin{figure}[ht!]
    \centering
    \includegraphics[width=\textwidth]{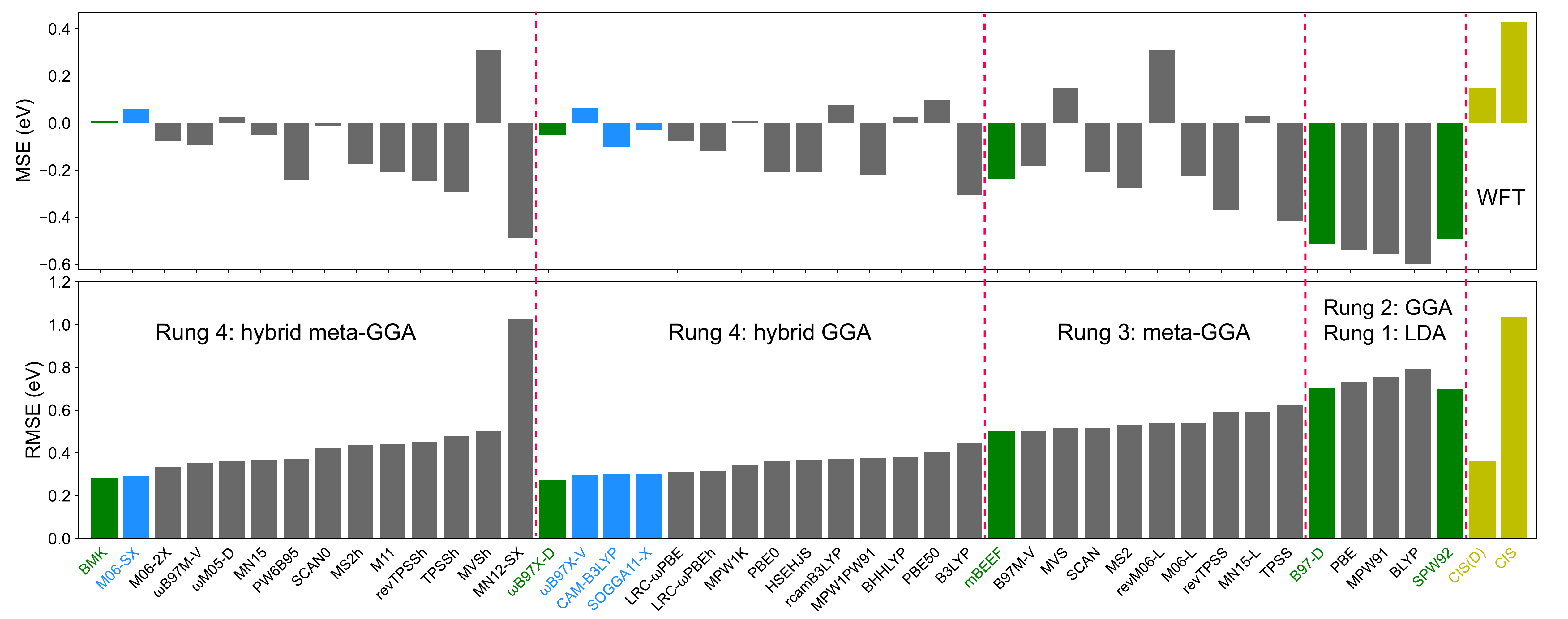}
    \caption{Mean signed erors (MSE) and root-mean square errors (RMSE) (both in eV) of 43 density functionals and two wavefunction methods (CIS, CIS(D)) on the entire benchmark dataset.}
    \label{fig:dft_barplot}
\end{figure}

Figure~\ref{fig:dft_barplot} and Figure~S1 summarizes the RMSE of 43 functionals on the whole benchmark dataset. N12, SOGGA11 and HFLYP are not considered here due to their very poor performance in the preliminary test on Q1. It turns out that Jacob’s ladder\cite{perdew2005prescription} is partially validated in the sense that mGGAs perform better than LDA and GGAs, while hybrid functionals perform better than local functionals. This indicates that the inclusion of additional physical content can indeed improve accuracy. Nevertheless, hybrid meta-GGAs have a much larger range of RMSEs than hybrid GGAs and the best hybrid meta-GGA could not outperform the best hybrid GGA, suggesting that the higher flexibility from the added physical content may not always have a positive effect on the prediction of excitation energies. This is consistent with studies on other molecular properties \cite{medvedev2017density, brorsen2017accuracy, hait2018dipole, hait2018polarizability, hait2021too} We expect that it is possible to train better hybrid meta-GGAs for predicting diverse molecular properties.

Diving a bit deeper into the data shown in Figure~\ref{fig:dft_barplot} and Table~S4.1 reveals a variety of other interesting observations:
\begin{itemize}[label={$\bullet$}]
    \item GGAs are not recommended for TDDFT calculation. The RMSE of all GGAs are higher than that of the SPW92 (i.e. rung 1). This is similar to the work of Leang et al.,\cite{leang2012benchmarking} wherein the SVWN LDA also surpassed all tested GGAs.
    \item Most functionals tend to underestimate excitation energies but many good functionals can attain a balance. Among local functionals, LDAs and GGAs yield strong underestimates  (MSE around -0.5 eV) while meta-GGAs can reduce the systematic error by roughly half. The addition of exact exchange (EXX) can further reduce systematic error and good RSH functionals (RSHs) or global hybrid functionals (GHs) with high EXX, like $\omega$B97X-D and BMK, can almost resolve this problem (the absolute values of their MSE are only near 0.05 eV).
    \item Taking the precision of the benchmark data ($\pm$ 0.05 eV) into consideration, the overall RMSE of good functionals on each rung are so close that we cannot give a definite answer about which functional is best for predicting excitation energies. $\omega$B97X-D offers the lowest overall RMSE (0.272 eV) among all functionals and BMK, M06-SX, $\omega$B97X-V, CAM-B3LYP and SOGGA11-X also provide similarly accurate predictions (RMSE $<$ 0.30 eV).  If semi-local functionals are required, we recommend mGGAs like mBEEF and B97M-V. They clearly outperform GGAs, although they are clearly inferior to the best hybrid functionals.
    \item TDDFT is highly recommended for electronic excitations that have dominant single-excitation character. All tested functionals outperform CIS and the best hybrid functionals can outperform CIS(D) even when we exclude the radical subset (QR), where CIS often gives qualitatively incorrect predictions and CIS(D) cannot provide satisfactory corrections.
\end{itemize}

\subsection{Performance for different excitation types}\label{subsec:types}

\begin{figure}[ht!]
    \centering
    \includegraphics[width=\textwidth]{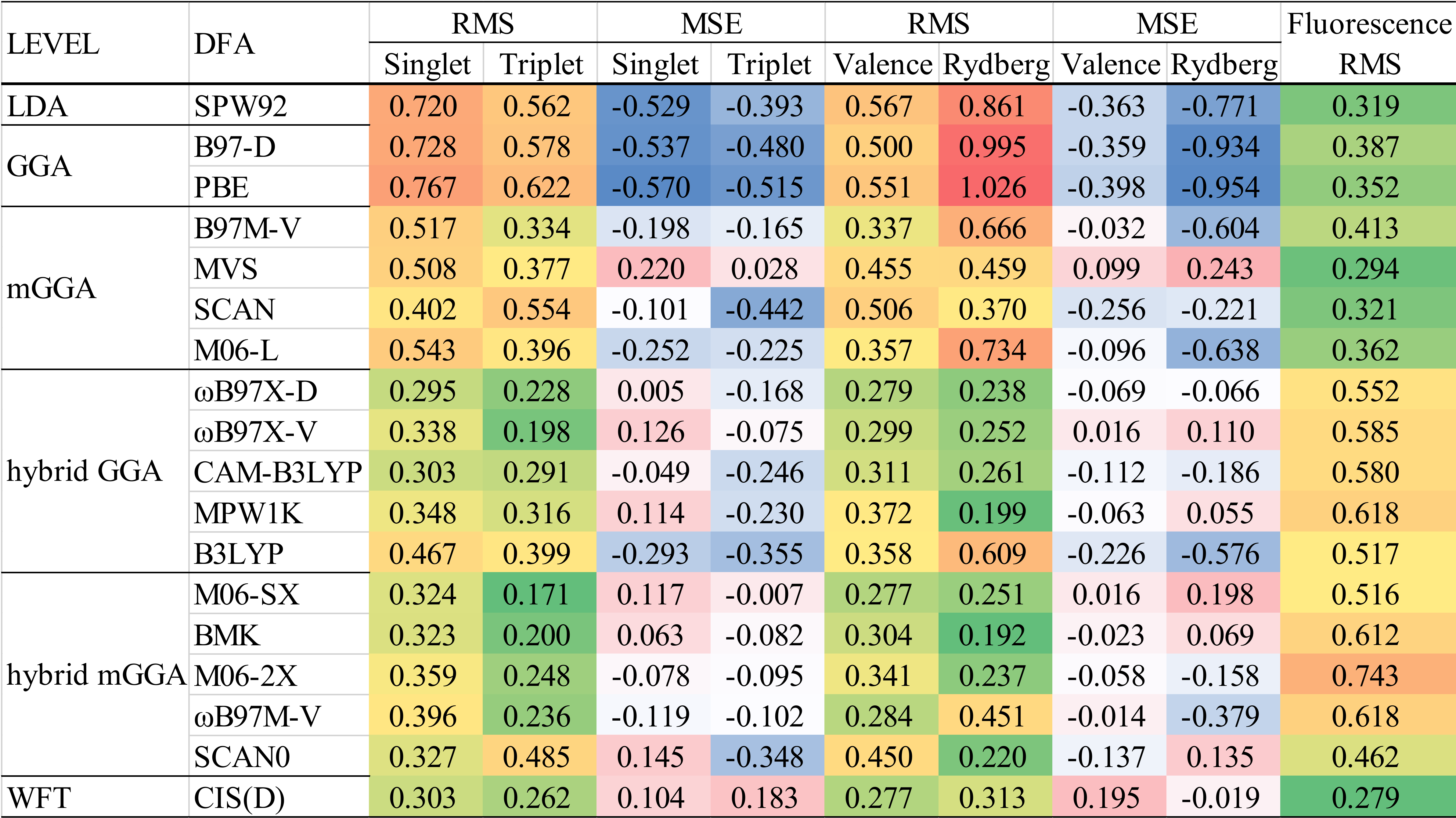}
    \caption{Comparison of RMSE and MSE (both in eV) of some good or popular functionals for the different types of excitation energies on the Q1, Q2 and QE dataset. RMSEs become smaller from red to green and MSEs become more positive from blue to red.}
    \label{fig:types}
\end{figure}

Figure~\ref{fig:types} shows some representative functionals' performance on different electronic excitation types---namely, singlet, triplet, valence, and Rydberg states. The comparison of all functionals can be found in Table~S4.7. We only use the Q1, Q2 and QE components of the dataset here because QCT and QR do not contain all classes of excitations. 

When we consider singlet and triplet excited states separately, it is evident that most functionals perform better for triplets than for singlets, which is consistent with the work of Leang et al.\cite{leang2012benchmarking} It also shows that hybrid functionals have more negative MSE for triplets because the triplet states have 2 parallel spins which are more stabilized by EXX.\cite{reiher2001reparameterization} We can see $\omega$B97X-D offers the best prediction for singlets (RMSE = 0.295 eV) and M06-SX offers the best prediction for triplets (RMSE = 0.171 eV).

\begin{figure}[htb!]
    \centering
    \includegraphics[width=\textwidth]{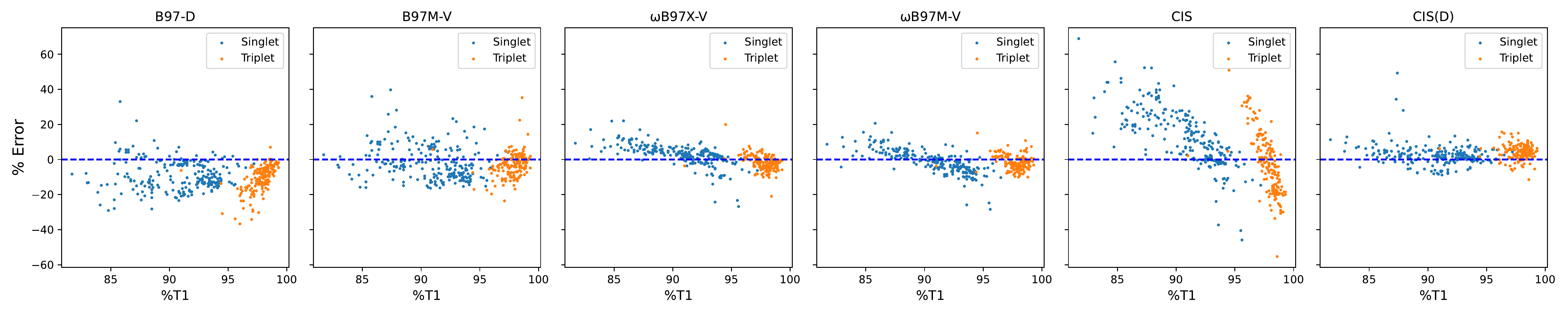}
    \caption{Comparison of percentage of relative error ($\%$ Error) and single excitation character computed at the LR-CC3 level.\cite{veril2021questdb} Singlets are shown in blue, and triplets in orange.}
    \label{fig:T1}
\end{figure}

The relation between error and single excitation character of the excited states was also studied for the triplet and singlet subsets. We expect that LR-TDDFT will overestimate excitation energies when the excited states have less single excitation character and their real energy is decreased by the multi-reference effect. However, Figure~\ref{fig:T1} shows that is only true for hybrid functionals and the CIS method. The CIS(D) method can partly solve the problem due to its double excitation terms. But the local functionals do not strongly show this trend either, suggesting that the performance of pure xc functionals may be essentially uncorrelated with extent of multi-reference character, at least in the scope we have chosen for our benchmark data  ($\%T_1 > 85$). 

For valence states, the better functionals of different rungs perform quite similarly. The best local functional, B97M-V, offers a RMSE of 0.337 eV and the best hybrid functional, M06-SX, only reduces the RMSE to 0.277 eV. In contrast, the functionals' RMSE differ greatly for Rydberg states. The LDAs and GGAs undervalue the Rydberg excitation energies a lot (RMSE $\approx$ 1.0 eV and MSE  $\approx$ -1.0 eV). Meta-GGAs can partly overcome this limitation of GGAs, and the best one, SCAN, can achieve a RMSE of 0.370 eV. The EXX term in good hybrid functionals, like BMK, MPW1K and $\omega$B97X-D, can reduce RMSE further to around 0.2 eV.

It deserves to be mentioned that the errors for calculated emission energies are surprisingly very different with those of absorption energies. From Figure~\ref{fig:types}, we see that CIS(D) offers a lower RMSE than all functionals and the local functionals demonstrates better performance than hybrid functionals. MVS, a mGGA, offers the best performance (RMSE = 0.294 eV) of the functionals, and even SPW92, the only LDA tested, offers the second lowest RMSE (0.319 eV) We also find the best functional on each rung is non-empirical. Therefore, one possible reason for poor performance is that the training sets of the semi-empirical functionals lack vertical emission energies, or even simply the ground state energies at non-equilibrium geometries. Another source of error is the effect of the ALDA in the LR-TDDFT method. Finally, given that there are only nine fluorescence states in the dataset, these conclusions should be read with caution.

\subsection{Performance for different benchmark datasets}\label{subsec:subsets}

\begin{figure}[htb!]
    \centering
    \includegraphics[width=0.85\textwidth]{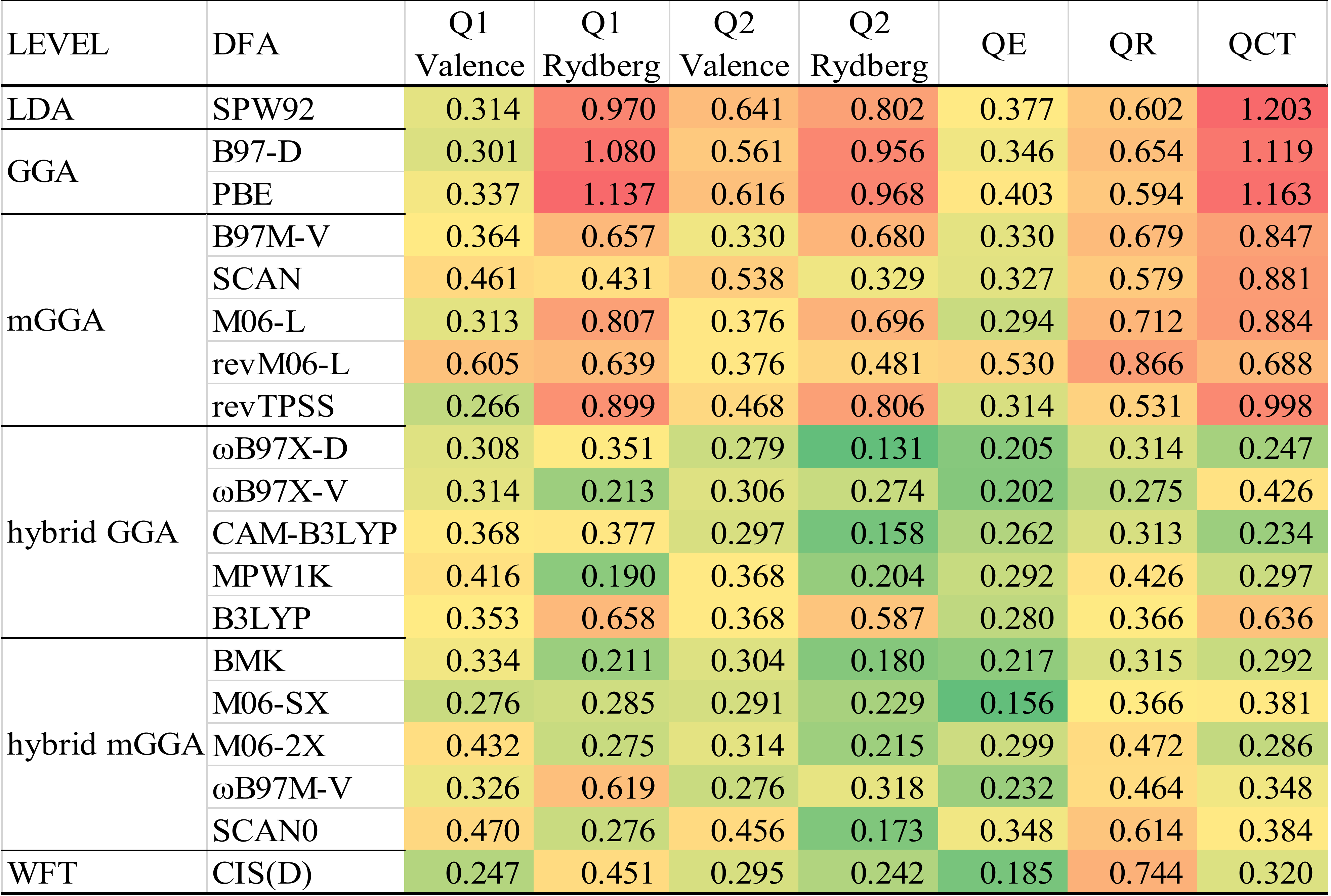}
    \caption{Comparison of RMSE (in eV) of some good or popular functionals for the excitation energies on different datasets. RMSE becomes smaller as the color changes from red to green in each column.}
    \label{fig:datasets}
\end{figure}

Figure~\ref{fig:datasets} shows the performance of some representative functionals on the different datasets. The comparison of all functionals can be found in Table~S4. 

The performances of functionals for valence and Rydberg states change a lot from Q1 dataset to Q2 dataset even though the difference between the two datasets is only the size of included molecules. For local functionals, on going from Q1 to Q2, the RMSE for Rydberg states decreases while the RMSE for valence states increases. One possible explanation is that the valence and Rydberg states mix more with each other for the bigger molecules in the Q2 dataset and their RMSE therefore become more similar. But this alone could not explain why the RMSE of both types decreases for most hybrid functionals. Another possibility is that the excitation energies of some molecules in Q1 dataset are more difficult to be predicted, such as those of nitrogen molecule.

There are another three special datasets, QE, QR and QCT, which corresponds to "exotic" molecules, radicals and intramolecular charge-transfer states, respectively. For the QE dataset, the RMSE of most methods is much smaller than that on Q1 or Q2 dataset, possibly because QE dataset only contains low-lying excited states of simple molecules and most of them are valence states. 

For the QR dataset, it is interesting that all the local functionals perform quite similarly. It implies that the inclusion of the electron density gradient and kinetic energy density is not useful for predicting the radical's excitation energies. By contrast, hybrid functionals can yield significantly improved performance with the help of EXX. It is worth mentioning that good hybrid functionals can provide much lower RMSE than CIS(D) for this dataset in the unrestricted formalism (we did not assess the restricted open shell variant\cite{head1995perturbative}). 

For the QCT dataset, the order of functionals is a little different than for the other datasets. Some functionals which perform very well for other datasets are less suitable for the QCT dataset, such as $\omega$B97X-V and M06-SX. $\omega$B97X-V fails possibly because it over-corrects most of the CT excitation energies, as shown by Figure~S7. The DFT hierarchy, however, is still kept. Similar to Rydberg states, the RMSE of mGGAs can be lower than LDA and GGA, and hybrid functionals, especially RSHs, and high EXX-containing GHs can do much better than the local functionals.

Overall, there is no local functional performing well for all datasets but there are some hybrid functionals which can perform very well for one dataset and are still not bad for other datasets. The best example seems to be BMK. Its RMSE is only a little higher than the best functional in each specific dataset and it thus becomes the best hybrid mGGA on the whole dataset. $\omega$B97X-D is another good example, and $\omega$B97X-V and M06-SX also have very good performance for all datasets other than QCT (for which they are more mediocre).

\section{Conclusions} \label{sec:conclusion}

(There are some minor changes in published paper.)

In this work, we have examined the performance of time-dependent density functional theory (TDDFT) for electronic excitations. This work builds on the many earlier assessments in several ways. First, we have taken the opportunity to assess promising functionals that have been developed over the past decade, and now widely used for ground state problems. For meta-GGA functionals, this required us to implement a correction for gauge invariance\cite{bates2012harnessing}. For functionals containing long-range dispersion, we implemented the appropriate response theory. Second, we based our assessment on the large set of high-quality benchmark calculations that comprise the QUEST database\cite{loos2018mountaineering, loos2020mountaineering1, loos2020mountaineering2, veril2021questdb, loos2021reference} produced over the past four years. This allowed our assessment to cover 463 separate electronic excitations. 

Our main conclusions are as follows:
\begin{enumerate}
    \item A careful assessment of the small molecule subset suggested that use of the Tamm-Dancoff approximation (TDA) is preferred over full TDDFT for excitations because of its improved accuracy versus cost tradeoff.
    \item Using the same small molecule subset, we demonstrate that excitation energies calculation using meta-GGAs do not typically benefit from the gauge invariance (GINV) correction. We do note that the M06-2X and BMK are two interesting exceptions, which exhibit significant improvement with use of GINV.
    \item The effect of non-local van der Waals density functionals on electronic excitations, while necessary for formal correctness of linear response theory, is shown to be very small.
    \item Across the full dataset, we find that TDDFT/TDA with the the best functionals yields RMSEs of 0.25-0.3 eV for excitation energies, which is a little better than the CIS(D) wavefunction method at a significantly lower computational cost.
    \item We recommend the BMK and $\omega$B97X-D functionals first because they can offer balanced predictions for different datasets and excitation types. M06-SX and $\omega$B97X-V also perform very well for most datasets, but have somewhat larger errors for QCT set. 
    \item Local functionals have higher error due to significant systematic underestimation of CT and Rydberg state energies, and we  cannot recommend GGAs. However two mGGAs, B97M-V and M06-L, turn out to be quite accurate for valence excitations.
    \item The Jacob's ladder hierarchy of functionals still partly exists for predicting vertical excitation energies since the best mGGA are better than LDA/GGAs, and hybrid functionals improve significantly over semi-local functionals. However, the LDA outperforms all the tested GGAs and hybrid mGGAs are not better than hybrid GGAs. Perhaps promising new functionals can still be obtained from the vast space of possible hybrid mGGAs\cite{Mardirossian:2016}.
\end{enumerate}
 
For the choice of basis sets, there are different recommendations for different purposes. For reaching the CBS limit, double augmentation with diffuse functions is required in nearly all circumstances. Only some good hybrid functionals can use single augmentation to approach the CBS limit of valence excitation energies. For practical purposes, we recommend aug-cc-pVDZ to predict valence states and def2-TZVPD or def2-QZVPPD to predict Rydberg states (with the expectation of some error cancellation between functional and basis set). 

In terms of caveats, we note that the datasets studied lack transition metal atoms and therefore our conclusions might only be applicable to organic/main group molecules. We also note that we have only assessed electronic excitations that are in the UV-vis region. TDDFT is known to perform much more poorly for core excitations, where other approaches such as state-specific orbital optimized DFT\cite{hait2020excited,hait2021orbital} appear to be preferable.\cite{hait2020highly} Indeed a full assessment of OO-DFT across the dataset used here would make for a very interesting comparison with TDDFT.

\begin{acknowledgement}
This work was supported by the Director, Office of Science, Office of Basic Energy Sciences, of the U.S. Department of Energy through the Gas Phase Chemical Physics Program, under Contract No. DE- AC02-05CH11231. Additional support to XF and MHG was provided through NIH grant 2R44GM121126-02. We thank Dr. Yuezhi Mao (Stanford) for providing valuable feedback on the implementation of VV10. JL would like to thank his wife Ming Sun for her help and companionship during this three-year project, especially during the difficult times of COVID-19.
\end{acknowledgement}

\section*{Supporting Information}

Additional figures (SI.pdf)

Raw data and Statistical errors (For readers of arxiv, please find these data in a folder called "anc" in source files. You can also download it from the published paper: 

https://doi.org/10.1021/acs.jctc.2c00160)

\textbf{Competing interests: } M.H-G. is a part owner of Q-Chem Inc., whose software was used in the calculations reported here. 

\clearpage

\bibliography{reference}
\end{document}